%% file: main.tex
\documentclass[sigconf,9pt,nonacm,balance=false]{acmart}
\usepackage{amsmath,array,tabularx}
\usepackage{placeins}
\usepackage{enumitem}
\usepackage[normalem]{ulem}

\usepackage{stfloats}
\usepackage{balance}
\usepackage[capitalise,nameinlink]{cleveref}
\crefname{appendix}{appendix}{appendices}
\Crefname{appendix}{Appendix}{Appendices}
\newcommand{\Shading}[1]{S_{\mathrm{#1}}}
\newcommand{\Layout}[1]{L_{\mathrm{#1}}}
\newcommand{\parasum}[1]{\textbf{#1}}
\newcommand{\Emboss}{Emboss}

\definecolor{rqbackground}{RGB}{232,241,248}
\newcounter{researchquestion}
\crefformat{researchquestion}{#2RQ#1#3}
\Crefformat{researchquestion}{#2RQ#1#3}
\crefname{researchquestion}{RQ}{RQs}
\Crefname{researchquestion}{RQ}{RQs}
\crefmultiformat{researchquestion}{#2RQ#1#3}{ and #2RQ#1#3}{, #2RQ#1#3}{, and #2RQ#1#3}
\Crefmultiformat{researchquestion}{#2RQ#1#3}{ and #2RQ#1#3}{, #2RQ#1#3}{, and #2RQ#1#3}
\newcommand{\rqbox}[1]{%
    \par\smallskip\noindent
    \colorbox{rqbackground}{%
        \parbox{\dimexpr\linewidth-2\fboxsep\relax}{%
            \raggedright\strut #1\strut}}%
    \par\smallskip
}
\newcommand{\researchquestion}[2]{%
    \refstepcounter{researchquestion}\label{#1}%
    \rqbox{\textbf{RQ\theresearchquestion.} #2}%
}
\newcommand{\rqanswer}[2]{\rqbox{\textbf{\Cref{#1}:} #2}}
\input{artifacts/evaluation_numbers}

\input{artifacts/production_numbers}

\input{artifacts/rq_numbers}
\input{artifacts/rq_synthesis_numbers}

\input{artifacts/rq_prediction_numbers}

\input{artifacts/rq_energy_accounting_numbers}

\title{Shading-Aware Rooftop PV Placement}
\author{Tobias von Arx}
\authornote{Also with University of Cambridge.}
\affiliation{%
    \institution{ETH Zurich}
    \city{Zurich}
    \country{Switzerland}
}
\email{tvonarx@ethz.ch}
\author{Srinivasan Keshav}
\affiliation{%
    \institution{University of Cambridge}
    \city{Cambridge}
    \country{United Kingdom}
}
\email{sk818@cam.ac.uk}
\hypersetup{pdftitle={Shading-Aware Rooftop PV Placement}}
\begin{document}
\begin{abstract}
Rooftop photovoltaic (PV) design requires careful panel placement to make effective use of potentially limited roof space and available sunlight.
Existing simplified roof models can miss objects and superstructures that shade panels.
We simulate rooftop irradiance using hourly weather and a terrain horizon, computing shadows from nearby buildings, trees, and roof superstructures rendered from the Sun's viewpoint.
Our method (i) infers existing panel layouts from imagery and estimates their yield, (ii) relocates these panels to improve the yield, and (iii) generates shading-aware layouts for new installations.
We compare modeled yield with production records from \MeasuredPVOutputCount{} Swiss installations and evaluate new layouts in simulation on \PatchRoofCount{} additional rural roofs.
Without knowledge of any electrical or system specifications, we obtain a median daily correlation with measured production of \EvalMedianDailyCorrelation{}.
Absolute energy estimates remain sensitive to system specifications.
Comparing new layouts with equal panel counts, accounting for neighborhood and roof-detail shading changes the position or orientation of an average of \PatchCombinedChangedMean\% of panels per roof.
These layouts increase modeled annual yield by up to \PatchCombinedGainMax\% compared with layouts designed without accounting for neighborhood and roof-detail shadows.
We publicly release our tool for modeling and designing rooftop PV installations with terrain, neighborhood, and roof-detail shading at \url{https://github.com/tobiasvonarx/shading-aware-pv}.
\end{abstract}
\ccsdesc[500]{Hardware~Renewable energy}
\ccsdesc[300]{Computing methodologies~Modeling and simulation}
\keywords{Rooftop photovoltaics, shading-aware panel placement, solar irradiance modeling, PV yield estimation, roof superstructures, shadow mapping}
\maketitle
\input{sections/introduction}
\input{sections/related_work}
\input{sections/method}
\input{sections/evaluation}
\input{sections/results}
\input{sections/discussion}
\input{sections/conclusion}
\clearpage
\nobalance
\bibliographystyle{ACM-Reference-Format}
\bibliography{references}
\appendix
\input{sections/appendix}
\end{document}

%% file: artifacts/evaluation_numbers.tex
\newcommand{\EvalMedianDailyCorrelation}{0.982}

\newcommand{\EvalMedianAbsoluteBias}{34.6}

%% file: artifacts/production_numbers.tex
\newcommand{\ProductionMinDays}{351}
\newcommand{\ProductionMedianSignedBias}{+34.6}
\newcommand{\ProductionMeanAbsoluteBias}{64.6}
\newcommand{\ProductionMeanBias}{60.1}
\newcommand{\ProductionMaxDays}{364}

%% file: artifacts/rq_numbers.tex
\newcommand{\MeasuredPVOutputCount}{11}

\newcommand{\FixedSeuzachNeighborhoodLoss}{2.14}
\newcommand{\FixedSeuzachDetailsLoss}{0.53}

\newcommand{\FixedMeasuredCombinedMedian}{0.31}
\newcommand{\FixedMeasuredCombinedMax}{2.67}
\newcommand{\FixedPatchNeighborhoodMedian}{6.22}

\newcommand{\FixedPatchDetailsMedian}{0.16}
\newcommand{\FixedPatchDetailsMax}{2.66}
\newcommand{\FixedPatchCombinedMedian}{6.22}
\newcommand{\FixedPatchCombinedMax}{18.81}
\newcommand{\DesignRelocationMedian}{0.22}
\newcommand{\DesignRelocationMax}{9.19}
\newcommand{\DesignRelocationPositive}{9}
\newcommand{\DesignRelocationFeasible}{11}

\newcommand{\DesignRelocationRetained}{2}
\newcommand{\DesignRelocationMaxHouse}{Grand-Lancy}

\newcommand{\PatchRoofCount}{19}
\newcommand{\ResultHalfMedianGain}{0.67}
\newcommand{\ResultHalfAboveOne}{9}
\newcommand{\ResultLocalUnchanged}{9}

\newcommand{\PatchNeighborhoodGainMean}{1.95}

\newcommand{\PatchCombinedGainMean}{2.03}
\newcommand{\PatchCombinedGainMax}{7.91}
\newcommand{\PatchCombinedChangedMean}{38.9}

\newcommand{\PatchLocalGainMax}{0.64}

\newcommand{\PatchExampleChanged}{3}
\newcommand{\PatchExampleGainKWh}{25.0}
\newcommand{\ResultLowMeanKWh}{157}
\newcommand{\ResultLowMedianGain}{1.10}
\newcommand{\BudgetLowGainMean}{2.98}
\newcommand{\ResultHalfMeanKWh}{322}

\newcommand{\ResultHighMeanKWh}{305}

\newcommand{\ResultFullMeanKWh}{34}
\newcommand{\ResultFullMedianGain}{0.11}
\newcommand{\BudgetFullGainMean}{0.13}
\newcommand{\ResultExampleAGain}{7.91}
\newcommand{\ResultExampleAKWh}{643.7}
\newcommand{\ResultExampleAChanged}{12}
\newcommand{\ResultExampleACount}{21}

\newcommand{\ResultExampleBGain}{3.89}
\newcommand{\ResultExampleBKWh}{147.7}
\newcommand{\ResultExampleBChanged}{8}
\newcommand{\ResultExampleBCount}{13}

\newcommand{\ResultExampleCGain}{0.27}
\newcommand{\ResultExampleCKWh}{26.4}
\newcommand{\ResultExampleCChanged}{2}
\newcommand{\ResultExampleCCount}{23}
\newcommand{\ResultExampleCShared}{93.9}
\newcommand{\DesignTriboltingenRelocation}{9.12}

%% file: artifacts/rq_synthesis_numbers.tex
\newcommand{\WithinMonthMedianR}{0.966}
\newcommand{\WithinMonthMinR}{0.790}
\newcommand{\WithinMonthMaxR}{0.974}

\newcommand{\MatchedBLoss}{18.81}

%% file: artifacts/rq_prediction_numbers.tex
\newcommand{\PredictionBaseMedianBias}{36.0}
\newcommand{\PredictionCapacityMedianBias}{25.7}
\newcommand{\PredictionShadingImproved}{10}
\newcommand{\PredictionCapacityImproved}{8}
\newcommand{\PredictionMedianAnnualizedMAEChange}{-52.7}

%% file: artifacts/rq_energy_accounting_numbers.tex
\newcommand{\EnergyBaselineNegative}{16}

%% file: sections/introduction.tex
\section{Introduction}

Residential rooftop photovoltaics (PV) are an important source of locally generated, renewable electricity~\citep{iea_pvps_trends_2025}.
Designing these installations involves making effective use of limited roof space and available sunlight.
However, chimneys, dormers, and other roof objects occupy space, preventing panel placement,
and can also shade nearby panels, greatly diminishing their output for certain hours of the day.
Surrounding buildings, trees, and distant terrain, such as hills, can also shade panels.
Therefore, when placing rooftop solar panels, it is important to carefully account for these impediments.

Existing methods model shadows from roof objects~\citep{krapf_deep_2022}, and other work uses irradiance estimates to guide panel placement~\citep{matsuoka_estimation_2024}.
However, shading losses alone do not establish how much yield can be gained by rearranging panels.
A shaded installation can benefit from rearrangement only if suitable alternative positions are available.
We therefore evaluate both the modeled losses at fixed panel positions and the gains from changing placement, for existing installations as well as new layouts on rural roofs.

To address this research gap, we present a method and publicly release a tool that infers existing panel layouts, estimates their yield, and proposes relocated or new layouts using terrain, neighborhood, and roof-detail shading.
We use \Emboss{}~\citep{vonarx_emboss_2026} to reconstruct roof geometry with superstructures and attachments, such as dormers, chimneys, vents, and skylights, which we summarize as \textit{roof details}.
We combine this geometry with light detection and ranging (LiDAR) measurements of the surroundings, the terrain horizon that describes which directions distant terrain blocks, and historical hourly weather data.
At each hour, we render the geometry from the Sun's viewpoint to determine which roof positions receive direct sunlight.
Without requiring (often unavailable) system-specific information, we then estimate electricity production using common assumptions about panel performance and the inverter, which converts the panels' direct current (DC) output to alternating current (AC).
Comparisons with measured production assess the accuracy of these estimates.
Holding weather and electrical assumptions fixed also lets us compare shading settings and panel layouts consistently within the model by considering the relative differences in yield.

We pose three research questions:

\begin{samepage}
\researchquestion{rq:yield}{How does the expected yield estimated from inferred layouts of existing installations compare with measured production?}
This establishes how well the model reproduces daily variation and total production when installation specifications are incomplete.
Comparing expected and measured yield can also help users troubleshoot their installations by highlighting unexpectedly low production.
\researchquestion{rq:shading}{How much yield is lost to neighborhood and roof-detail shading for a given panel layout?}
Holding panel positions fixed isolates the effect of each shading source on modeled yield.
\researchquestion{rq:design}{How does shading-aware placement affect panel layouts and modeled yield?}
We then allow panel positions to change, while keeping panel counts fixed, to determine how much modeled yield can be gained through relocation or shading-aware design.
\end{samepage}

To evaluate our approach, we compute
yield estimates, shading losses, and panel relocation on \MeasuredPVOutputCount{} measured installations, and new-layout design on \PatchRoofCount{} additional roofs in a Swiss rural patch.
We find that expected yield follows measured daily variation closely, with a median correlation of \EvalMedianDailyCorrelation{}.
Without knowledge of any electrical or system specifications, the yield estimates have a median absolute energy bias of \EvalMedianAbsoluteBias\% relative to measured production.
Using the reported DC capacity reduces this bias to \PredictionCapacityMedianBias\%, while inverter characteristics, wiring, system age, and other factors remain unknown.
Relocation improves modeled yield on \DesignRelocationPositive{} of \MeasuredPVOutputCount{} inferred installations.
When generating new layouts for rural roofs, sized at half the maximum feasible panel capacity on each roof, optimized layouts that account for neighborhood and roof-detail shading move or rotate an average of \PatchCombinedChangedMean\% of panels per roof and increase modeled annual yield by up to \PatchCombinedGainMax\% over layouts that omit those shadows.
Benefits vary across roofs, highlighting the value of assessing shading and feasible placement together.

Our main contributions are:
\begin{itemize}
    \item A method to infer panel layouts in existing installations using remote sensing
    \item A public tool that combines terrain, neighborhood, and roof-detail shading to estimate yield, evaluate panel relocation, and design new layouts.
    \item An evaluation using production records from \MeasuredPVOutputCount{} installations and simulations on \PatchRoofCount{} additional rural roofs, distinguishing yield agreement, shading losses, and placement gains.

\end{itemize}

\Cref{sec:related-work} positions these tasks within prior work.
\Cref{sec:methods} describes the modeling and placement procedure, and \Cref{sec:evaluation} defines the comparisons used to answer each research question.
\Cref{sec:results} presents the answers in the same order, followed by their implications and limitations in \Cref{sec:discussion}.

%% file: sections/related_work.tex
\section{Related work}
\label{sec:related-work}

We outline prior work on PV production estimation, roof geometry, shading, and panel placement.
\Cref{tab:related-capabilities} compares the capabilities of prior approaches.

\begin{table}[!htbp]
    \caption{Shading and layout capabilities of selected PV assessment and placement approaches.
    \ding{51}: supported, \ding{55}: not supported as per our definitions, ?: unknown.}
    \label{tab:related-capabilities}
    \centering
    \begingroup
    \small
    \setlength{\tabcolsep}{4pt}
    \renewcommand{\arraystretch}{1.08}
    \newcommand{\caphead}[1]{\rotatebox{90}{\strut #1}}
    \newcommand{\yes}{\ding{51}}
    \newcommand{\no}{\ding{55}}
    \begin{tabular*}{\columnwidth}{@{\extracolsep{\fill}}lcccccc@{}}
        \toprule
        & \multicolumn{3}{c}{Shading} & \multicolumn{3}{c}{Layout} \\
        \cmidrule(lr){2-4}\cmidrule(l){5-7}
        Method
        & \caphead{Terrain}
        & \caphead{Neighborhood}
        & \caphead{Roof details}
        & \caphead{\shortstack[l]{Existing-layout\\inference}}
        & \caphead{Panel placement}
        & \caphead{\shortstack[l]{Shading-aware\\placement}} \\
        \midrule
        PVGIS 5.3~\citep{pvgis_api} & \yes & \no & \no & \no & \no & \no \\
        Google Solar~\citep{google_solar_api_building_insights_2026,google_solar_methodology_2026} & \yes & \yes & \yes\textsuperscript{*} & \no & \yes & \yes \\
        SolarNet+~\citep{li_deep_2024} & \yes & \no & \no & \no & \yes & \no \\
        Krapf et al.~\citep{krapf_deep_2022} & \yes & \yes & \yes & \no & \yes & \no \\
        Matsuoka et al.~\citep{matsuoka_estimation_2024} & \no & \yes & \no & \no & \yes & \yes \\
        Tian et al.~\citep{tian_semantic_2025} & ? & \yes & \yes & \no & \no & \no \\
        Wang et al.~\citep{wang_shading_2026} & \yes & \yes & \no & \no & \no & \no \\
        Awwad et al.~\citep{awwad_site_2023} & \no & \no & \no & \no & \yes & \yes \\
        Miao et al.~\citep{miao_layout_2026} & \no & \yes & ? & \no & \yes & \yes \\
        SunPlace~\citep{gschwind_joint_2026} & \yes & \yes & \yes\textsuperscript{*} & \no & \yes & \yes \\
        \midrule
        \textbf{Ours} & \yes & \yes & \yes & \yes & \yes & \yes \\
        \bottomrule
    \end{tabular*}
    \par\smallskip
    \parbox{\columnwidth}{\footnotesize\raggedright
    * Only larger details are modeled.\par
    \textbf{Capability definitions}: For roof details, we require explicit shading modeling, beyond exclusions. Existing-layout inference means recovering individual panel positions, which are needed to evaluate relocation.}
    \endgroup
\end{table}

\parasum{Solar resource and production estimates.}
PVGIS provides irradiance and PV time series from satellite and reanalysis weather, with terrain-horizon corrections~\citep{pvgis_api}.
\citet{pfenninger_pv_2016} compare hourly simulations with records from over 1,000 PV systems and national production data, examining temporal variation and systematic bias.
These comparisons address how well weather-driven simulations reproduce production, which also depends on installation-specific electrical parameters.

\parasum{Roof geometry and usable area.}
Estimating how much electricity a roof could generate requires knowing how many panels can fit and their tilt and orientation. Aerial imagery can help identify usable roof area and estimate roof geometry.
SolarMTNet jointly infers roof orientations and slopes to estimate rooftop irradiance, without modeling superstructures~\citep{boccalatte_multitask_2025}.
SolarNet+ detects superstructures and packs panels into the remaining roof area, but omits neighborhood shadows~\citep{li_deep_2024}.
\citet{boccalatte_superstructures_2025} exclude roof segments with detected superstructures and quantify the loss of usable area and solar potential, without considering panel placement or shadowing.
\citet{krapf_deep_2022} add image-derived roof details to 3D city models and evaluate their shading- and obstruction effect specifically in fully-packed layouts.
Their study finds that roof details change estimated irradiation (using the engine of \citet{willenborg_city_2018}) significantly through limiting usable area.
It leaves open how yield can be gained by using those shadows to optimize the placement of PV panels.
Our evaluation uses these shadows during placement to assess changes in layouts and modeled yield.
Note that SolarNet+ and Krapf et al. do segment existing PV regions, but do not recover individual panels.

\parasum{Spatially-resolved shading.}
Shading models estimate how sunlight varies across usable roof area.
\citet{matsuoka_estimation_2024} determine sunlight visibility by rendering buildings from the Sun's viewpoint, avoiding ray tracing from each roof position. They use the resulting hourly irradiance to guide panel placement, but do not infer or relocate existing panels, or consider roof details.
We adopt their rendering approach for our shadow simulation.
\citet{tian_semantic_2025} classify obstructions in urban point clouds and separate shading contributions from different object types, but leave detailed panel arrangement to later design stages.
\citet{wang_shading_2026} assess city-wide irradiance on a 2\,m surface grid using terrain and neighborhood horizons, without considering panel layouts.

\parasum{Shading-aware placement and sizing.}
Shading-aware placement methods choose feasible panel positions using estimates of the sunlight available at each position. Methods differ in whether they maximize energy yield or cost economics.
\citet{zhong_spatial_2022} combine roof geometry, panel dimensions, and irradiance in a maximum-cover model for new layouts, without inferring existing panel arrangements.
\citet{barbon_general_2022} optimize panel rows on irregular flat roofs, leaving roof detail shadows unmodeled, and \citet{awwad_site_2023} use mixed-integer nonlinear programming to optimize panel positions, azimuths, and tilts with mutual panel shading, focusing on new installations with adjustable panel orientations.
\citet{alharbi_economical_2023} jointly optimize system size and panel orientations for net present value, accounting for self-shading, roof access, and inverter type.
Our study, instead, examines how neighborhood and roof-detail shading affect inferred installations, relocation gains, and new layouts.

At the building-cluster scale, \citet{miao_layout_2026} maximize panel count within each roof's available area, then select which buildings receive these layouts to balance energy, economic, and carbon objectives.

Other integrated tools connect spatial assessment directly to installation planning.
Google Solar API combines imagery-derived geometry, surrounding shadows, and historical weather to return panel proposals and energy estimates~\citep{google_solar_api_building_insights_2026,google_solar_methodology_2026}.
SunPlace combines hourly ray tracing and greedy non-overlapping placement with PV and battery sizing and electric-vehicle operation~\citep{gschwind_joint_2026}.
Its roof-detail shading is limited to larger features represented in the input geometry.
Our placement follows its ranking principle, scoring candidate positions by annual irradiation.

\parasum{Our contribution.}
We combine reconstruction, shading, and placement methods in one workflow that infers individual installed panels, estimates their yield, evaluates relocation, and generates new layouts.
This lets us compare yield estimates with measured production and distinguish the energy lost to shading at fixed positions from the energy gained by changing those positions.

%% file: sections/method.tex
\section{Method}
\label{sec:methods}

\begin{figure*}[!tbp]
    \centering
    \input{figures/method_overview_values}
    \includegraphics[width=0.86\textwidth]{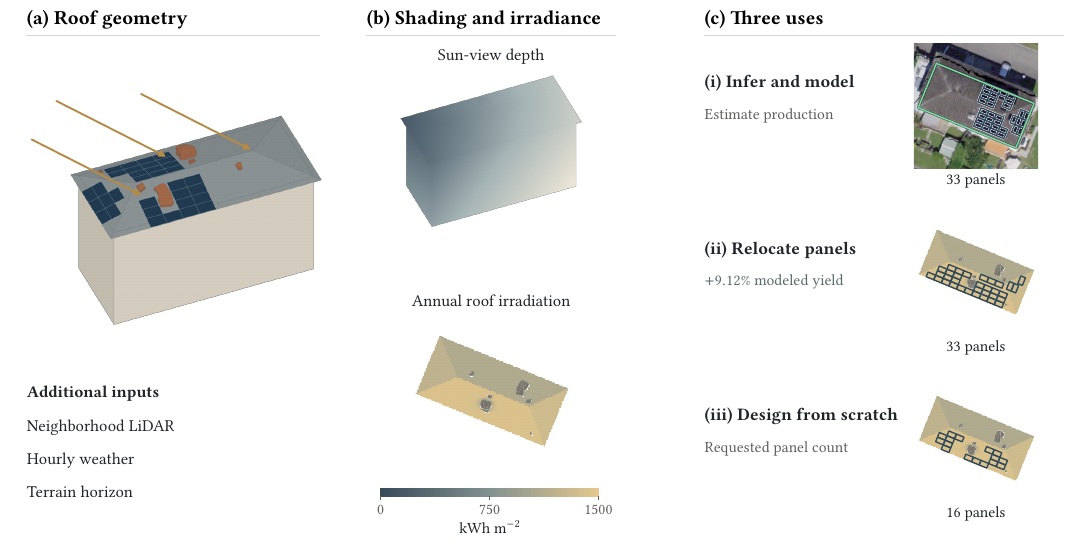}
    \caption{Our shading model, illustrated with a house in Triboltingen, Switzerland.
    (a) Reconstructed roof geometry and inferred panels, with parallel rays following the Sun direction, and additional inputs.
    (b) Surface depth along the Sun direction on \MethodSunSnapshot{}, shown from the same viewing angle as (a), with darker tones nearer to the Sun. The neighborhood uses a second Sun-view depth buffer, not shown here.
    The bottom panel shows annual solar energy per unit area received in the panel plane at panel height, projected onto the roof.
    (c) Imagery-inferred panels, their same-count relocation, and a new layout with 16 panels.}
    \Description{Reconstructed roof geometry, neighborhood LiDAR, hourly weather, and a terrain horizon are inputs to our shading model. Visibility and terrain-corrected hourly weather determine irradiance across the roof. The three branches show imagery-inferred panels for expected-yield estimation, relocation at the inferred count with its modeled yield gain, and a new layout at a requested count. Warm colors show annual irradiation.}
    \label{fig:method-overview}
\end{figure*}

We combine a detailed roof model, orthoimagery, neighborhood LiDAR, terrain-corrected irradiance, and hourly temperature and wind data to estimate roof irradiance (\Cref{sec:shadow-states}). Annual irradiation guides placement, and an electrical model estimates AC yield (\Cref{sec:placement-yield}).
Our workflow evaluates inferred installations, relocates their panels at fixed count, and designs new layouts (\Cref{fig:method-overview}).

\subsection{Computing irradiance on the roof}
\label{sec:shadow-states}

We account for shading at three spatial horizons: distant terrain, nearby buildings and vegetation, and the roof itself.
For terrain, we use the Photovoltaic Geographical Information System (PVGIS), version~5.3, with its satellite-derived SARAH3 irradiance dataset and associated temperature and wind data~\citep{pvgis_api}.
Its terrain-horizon correction accounts for sunlight blocked by distant topography, such as hills.

To model shading from nearby buildings and vegetation, we use LiDAR point cloud data within a 200~m square centered on the roof.
We divide this area into a horizontal grid of $1\times1$~m cells and assign each cell the highest LiDAR elevation within it.
If a cell contains no points, we find the nearest cell with LiDAR points classified as ground and use the highest of those ground elevations.
We connect adjacent cell centers into triangles to form the surface rendered from the Sun's viewpoint.
We model vegetation as being opaque to direct sunlight, and do not model seasonal canopy changes.

To model roof-detail shading, we use \Emboss{}~\citep{vonarx_emboss_2026} to add roof details to the national swissBUILDINGS3D building models.
\Emboss{} does so using SWISSIMAGE aerial imagery and swissSURFACE3D LiDAR measurements~\citep{swisstopo_geodata}.
For a target building, this detailed roof model replaces its LiDAR representation, allowing us to evaluate roof and neighborhood shading separately.
We assume panels lie parallel and close to the roof surface (flush mounting), and omit shadows cast by the panels themselves.

With the geometry assembled, we sample sunlight at points called \textit{receivers}, placed just above the roof to represent panel height.
The points follow a regular horizontal grid. Each is weighted by the corresponding area on the sloped roof, so steeper segments receive the appropriate area weight.
\Cref{app:settings} gives the grid spacing and height offset.
We compute the Sun's position and transform the weather irradiance to each roof's orientation using the pvlib software library~\citep{holmgren_pvlib_2018}.
This transformation includes the Perez model for diffuse sunlight scattered by the sky onto a tilted surface~\citep{perez_modeling_1990}.

Adapting \citet{matsuoka_estimation_2024}, at each daylight timestep we render the detailed roof and the LiDAR-derived neighborhood separately from the Sun's viewpoint.
For each scene, the depth buffer stores the depth of the nearest surface at each pixel.
We project each receiver into this view and compare its depth along the viewing direction with the stored depth at that pixel.
The receiver is shaded if it lies behind the stored surface.
Under full shading, a receiver receives direct sunlight only if neither scene blocks it.
Terrain shading is accounted for through a terrain-horizon correction done by PVGIS.

Combining the weather data with this visibility test gives the irradiance incident on the panel plane, called plane-of-array (POA) irradiance. For receiver $i$ and hour $t$, it is
\begin{equation}
G_{it}=v_{it}G^{\mathrm{beam}}_{it}+G^{\mathrm{sky}}_{it}+G^{\mathrm{ground}}_{it}\quad[\mathrm{W\,m^{-2}}]
\label{eq:irradiance}
\end{equation}
where $v_{it}=1$ when direct sunlight reaches the receiver and $v_{it}=0$ when the modeled geometry blocks it.
The three terms represent direct sunlight, diffuse sky light, and ground-reflected light, respectively.
Shadows reduce only the direct term in this model. We do not model obstruction of diffuse or ground-reflected light.

To separate the effects of different shadow sources, allowing us to quantify yield losses from neighborhood and roof-detail shading separately, we evaluate four shading settings:
\begin{itemize}[leftmargin=*,labelindent=0pt]
    \item $\Shading{Base}$ \hfill terrain and building self-shading
    \item $\Shading{Neighborhood}$ \hfill Base plus nearby buildings and trees
    \item $\Shading{RoofDetails}$ \hfill Base plus roof-detail shadows
    \item $\Shading{Full}$ \hfill Base plus both additions
\end{itemize}

Here, building self-shading comes from the main roof and building body, excluding added roof details. For example, a higher roof section can shade a lower one. We call this, together with terrain shading, base-roof shading.

These settings change the shadows used to estimate sunlight.
Roof details remain physical obstructions to panel placement in every setting, even when their shadows are omitted.
Our released tool uses $\Shading{Full}$, which includes all modeled shadow sources.

\subsection[Selecting panel positions and estimating yield]{Selecting panel positions and estimating yield}
\label{sec:placement-yield}

A layout specifies panel positions and orientations.
We infer it from imagery for existing installations or generate it for relocation and new designs.
For existing installations, the \Emboss{} segmentation model identifies image regions containing PV panels.
We fit rectangular grids of panels, parallel to the roof surface, to these regions.
We test panel dimensions and select the grid that best matches the segmented area and visible panel boundaries, favoring common panel sizes (\Cref{app:settings}).
All panels on a house use the same dimensions. Fitting uses neither system specifications nor known panel counts.
The inferred panel footprints determine where we sample irradiance to estimate production.

Relocation preserves inferred panel counts and dimensions. New layouts use requested counts and dimensions. In both cases, \textit{feasible} placements keep panels within roof boundaries and satisfy obstruction clearances (\Cref{app:settings}).

We construct candidate positions separately on each planar roof segment.
For each segment, we test grids in portrait and landscape orientation, shifting each grid across the usable area in regular steps (\Cref{app:settings}).
We retain the grid that fits the most panels. All panels in this grid share an orientation, but different roof segments can use different orientations.
If several grids fit the same number, we choose the one with the largest sum of annual irradiation scores beneath its panels.

We construct a set $\mathcal{C}$ of possible panel placements from the grids selected for each roof segment.
Each placement, called a candidate, specifies the location and orientation of one panel.
For relocation, we also include the inferred placements of existing panels.

For each candidate $j$, we select all receiver points (\Cref{sec:shadow-states}) on the same roof segment that lie within the panel's rectangular footprint.
Its score $q_j$ is the area-weighted mean annual plane-of-array irradiation at those receivers [$\mathrm{kWh\,m^{-2}}$].
We denote by $\mathcal{O}$ the pairs of candidates whose panel footprints overlap and therefore cannot both be selected.
The generated grids contain no overlapping panels, so $\mathcal{O}$ is empty for new layouts.
For relocation, overlaps can arise between generated and inferred placements.
To select a layout of $N$ panels with the highest total irradiation score and no overlapping footprints, we solve
\begin{equation}
\begin{aligned}
\max_{\{x_j\}} \quad & \sum_{j\in\mathcal{C}} q_j x_j\\
\text{subject to}\quad
& \sum_{j\in\mathcal{C}} x_j=N\\
& x_j+x_k\leq 1 && \forall (j,k)\in\mathcal{O}\\
& x_j\in\{0,1\} && \forall j\in\mathcal{C}
\end{aligned}
\label{eq:placement}
\end{equation}
Here, $x_j=1$ selects position $j$.
For new layouts, this reduces to choosing the $N$ highest scores, following the ranking principle of SunPlace~\citep{gschwind_joint_2026}.

We estimate panel temperature from area-weighted hourly POA irradiance, air temperature, and wind speed using the Faiman model~\citep{faiman_temperature_2008}.
We then estimate AC energy using temperature-dependent DC conversion, a fixed inverter efficiency, and an inverter power limit, with component assumptions and details specified in \Cref{app:electrical}.
Summing hourly AC energy gives the modeled annual yield.

The placement objective maximizes annual irradiation, but the final comparison concerns AC energy.
Higher irradiation need not give higher AC yield, because warmer panels convert sunlight less efficiently and output above the inverter's power limit is lost through clipping.
We therefore evaluate every proposed layout with the electrical model.
For relocation, we select a proposal using irradiation under full shading and retain it only if it increases modeled AC energy over the inferred layout.
Relocation gains are evaluated in simulation, with production measurements available only for the existing installation.
Note that our estimates deliberately avoid installation-specific metadata, so differences in installed capacity, degradation, wiring losses, and operating conditions can affect their accuracy (\mbox{\Cref{sec:rq1-results}}).

Our web interface lets users select Swiss buildings, inspect inferred installations and modeled yield, and explore relocated or new layouts (\Cref{app:interface}).

%% file: figures/method_overview_values.tex
\newcommand{\MethodSunSnapshot}{21 March 2023 at 09:10 UTC}

%% file: sections/evaluation.tex
\section{Data and Evaluation Approach}
\label{sec:evaluation}

We evaluate production agreement, shading losses at fixed panel positions, and gains from changing layouts (\Cref{rq:yield,rq:shading,rq:design}).

\begin{figure}[!htbp]
    \centering
    \includegraphics[width=\linewidth]{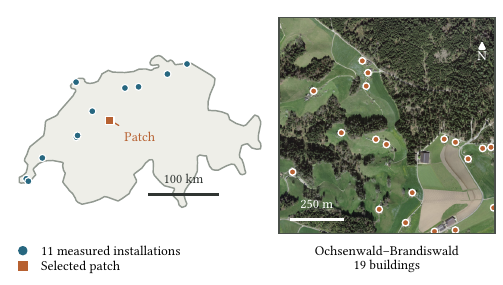}
    \caption{Study locations. Left: \MeasuredPVOutputCount{} measured installations in Switzerland, with the rural study area marked in orange. Right: the $1\times1$~km Ochsenwald--Brandiswald patch and its \PatchRoofCount{} residential buildings.}
    \Description{A map of Switzerland locates eleven measured PV installations and the selected rural patch. An adjacent aerial view shows the one-kilometre square with dots at the nineteen residential building coordinates.}
    \label{fig:data-settings}
\end{figure}

\Cref{fig:data-settings} shows the two groups of houses in our case study.
For installations with publicly available production data, we compare estimated and measured yield and model the gains from relocating panels.

Existing PV systems, especially those with publicly shared data, are usually installed by solar enthusiasts, who take care to keep their panels unshaded.
Their PV layouts are therefore less likely to be placed in the shade of roof- or neighborhood-details.
To reduce this bias, we also study a typical $1 \times 1$ km rural patch, which lets us examine novel layouts and the effects of different shading sources.

\parasum{Measured installations.}
We manually select \MeasuredPVOutputCount{} installations from PVOutput~\citep{pvoutput_api} and evaluate each over one year.
Across the houses, \ProductionMinDays{}--\ProductionMaxDays{} days per installation have reported generation.
As described in \Cref{sec:methods}, we infer their PV layouts from imagery and compare expected yield with measurements (\Cref{rq:yield}).

\parasum{Rural patch.}
We study all \PatchRoofCount{} buildings with full or partial registered residential use whose coordinates fall within the selected square.
These roofs have no production records, so we evaluate only modeled shading effects and new layouts (\Cref{rq:shading,rq:design}).

\Cref{fig:house-atlas} in \Cref{app:house-atlas} provides imagery and a layout overview of every evaluated roof.

\parasum{\Cref{rq:yield}: compare expected yield with measurements.}
We use weather data from the measurement year and compare modeled and measured daily energy on days with complete data for both series.
Daily correlation measures whether predicted production rises and falls with the measurements. Energy bias measures the difference in their totals, divided by measured energy over those same days.
The production comparison includes modeled snow losses where the required station data are available (\Cref{app:settings}).
We first use default electrical parameters, then repeat the comparison with reported DC capacity to assess the effect of that assumption.

\parasum{\Cref{rq:shading}: evaluate the same layout under different shadow sources.}

We now compare simulations to isolate shading effects.
For both this analysis and \Cref{rq:design}, we omit snow losses and keep weather and electrical assumptions the same within each comparison.
To measure shading losses without changing placement, we hold the inferred layouts fixed on the \MeasuredPVOutputCount{} measured installations and use layouts designed under base-roof shading at 50\% of the feasible panel count on the \PatchRoofCount{} rural roofs.
Starting from the base setting, which includes terrain and shadows cast by the target building's base geometry, we add neighborhood shadows and then roof-detail shadows.
Both energy reductions are divided by the energy in the base setting, so their percentages add to the total loss.
Because shadows can overlap, the roof-detail increment measures the extra loss after neighborhood shadows are already present.
Writing $E(L,S)$ for annual modeled AC energy [kWh] of layout $L$ under shading setting $S$, the loss under full shading relative to base-roof shading is
\begin{equation}
\ell(L)=1-\frac{E(L,\Shading{Full})}{E(L,\Shading{Base})}
\label{eq:fixed-loss}
\end{equation}

\parasum{\Cref{rq:design}: evaluate relocation and shading-aware design.}
On the \MeasuredPVOutputCount{} measured installations, we compare inferred and relocated layouts under full shading, preserving panel counts and dimensions. Gains may come from avoiding shade or using better-oriented roof segments. The original installation's design criteria are assumed to be unknown.

On the \PatchRoofCount{} rural roofs, we assess the benefit of accounting for shading during design by comparing layouts generated under base-roof, neighborhood, and full shading.
All designs use the same placement procedure, usable roof geometry, $1.0\times1.7$~m panels, and clearances.
For each roof, we first determine the feasible panel count under each of the four shading settings: base-roof, neighborhood, roof-detail, and full shading.
Each setting selects grids that fit the most panels, using annual irradiation to break ties.
We use the smallest of the four resulting counts as the common feasible panel count and generate layouts at 20\%, 50\%, 80\%, and 100\% of this count.
We evaluate all layouts under full shading to test whether including more shadow sources during design improves modeled yield.
For layouts $\Layout{Base}$ and $\Layout{Full}$ designed under base-roof and full shading, respectively, the gain is
\begin{equation}
g=\frac{E(\Layout{Full},\Shading{Full})}{E(\Layout{Base},\Shading{Full})} - 1
\label{eq:design-gain}
\end{equation}
\begin{sloppypar}
Relocation replaces $\Layout{Base}$ in the above expression with the inferred layout.
The neighborhood gain compares $\Layout{Neighborhood}$ with $\Layout{Base}$, normalized by the full-shading energy of $\Layout{Base}$.
The roof-detail gain compares $\Layout{Full}$ with $\Layout{Neighborhood}$, normalized by the full-shading energy of $\Layout{Neighborhood}$.
These percentage gains use different denominators and therefore do not add to the full-versus-base percentage gain.
\end{sloppypar}

We report relative gains as percentages and annual energy differences in kWh.
To measure layout changes, we pair panels between layouts to minimize the sum of distances between their 3D centers, using each panel once.
We count a panel as changed when its displacement or rotation exceeds the tolerances in \Cref{app:settings}, so small differences do not count as layout changes.
For rural designs, we also ask why changing positions improves yield.
A new position may avoid neighborhood or roof-detail shadows while receiving less sunlight because of its roof orientation.
To distinguish these effects, we separate the energy change under base shading from the reduction in additional shading loss.
We obtain them by evaluating $\Layout{Base}$ and $\Layout{Full}$ under both $\Shading{Base}$ and $\Shading{Full}$.
Define the additional shading penalty as $P(L)=E(L,\Shading{Base})-E(L,\Shading{Full})$.
The gain in annual energy then separates into
\begin{equation}
\begin{aligned}
\Delta E &= E(\Layout{Full},\Shading{Full})-E(\Layout{Base},\Shading{Full})\\
         &= E(\Layout{Full},\Shading{Base})-E(\Layout{Base},\Shading{Base})\\
         &\quad +P(\Layout{Base})-P(\Layout{Full})
\end{aligned}
\label{eq:energy-accounting}
\end{equation}
The first difference measures the energy change from repositioning panels under base-roof shading.
The second measures the neighborhood and roof-detail shading loss avoided by the new layout.

%% file: sections/results.tex
\begin{figure*}[!b]
    \centering
    \includegraphics[width=0.86\linewidth]{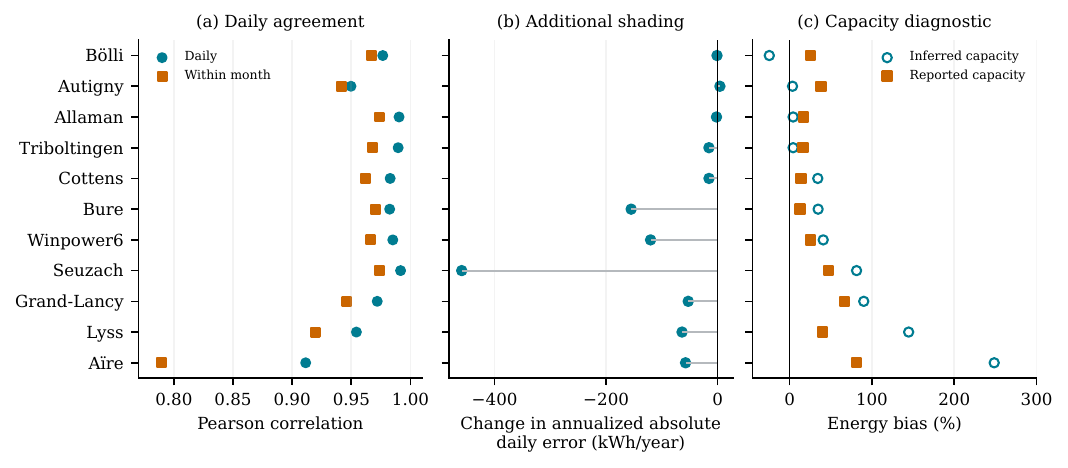}
    \caption{Expected yield versus measured production and two diagnostics on all \MeasuredPVOutputCount{} PVOutput systems, ordered by increasing bias.
    (a) Daily Pearson correlation, before and after subtracting each series' own monthly mean.
    (b) Change in mean absolute daily error (MAE), multiplied by 365 to express it as an annualized value, when neighborhood and roof-detail shadows are added to base-roof shading. Negative values indicate improvement.
    (c) Bias under full shading using DC capacity estimated from inferred panel area versus reported DC capacity, preserving the original DC/AC ratio.
    Bias is the difference between modeled and measured energy, normalized by measured energy.}
    \Description{Three panels distinguish temporal agreement, the small change from adding shading, and the larger effect of replacing inferred capacity with reported capacity. Substantial energy bias remains.}
    \label{fig:yield-validation}
\end{figure*}

\section{Results}
\label{sec:results}

We report agreement with measured production for \Cref{rq:yield}, fixed-layout shading losses for \Cref{rq:shading}, and modeled placement gains for \Cref{rq:design}.
This separates the effect of shading on an installation from the benefit of using shading information during placement.

\subsection{Expected yield for existing installations}
\label{sec:rq1-results}

We compare expected yield with daily production records for the \MeasuredPVOutputCount{} measured installations.
We first assess daily variation and total energy, then examine whether shading and capacity assumptions explain the discrepancies.

\begin{table}[!htbp]
    \centering\small
    \setlength{\tabcolsep}{1.2pt}
    \caption{Expected yields for all \MeasuredPVOutputCount{} PVOutput systems.
    Panel counts are inferred/reported. Inferred counts represent the segmented PV area using estimated panel dimensions and can differ from reported counts (\Cref{app:house-atlas}).
    Bias is relative to measured energy, and $r$ is daily Pearson correlation.}
    \label{tab:production}
    \input{artifacts/production_table}
\end{table}

\begin{figure}[!htbp]
    \centering
    \includegraphics[width=\linewidth]{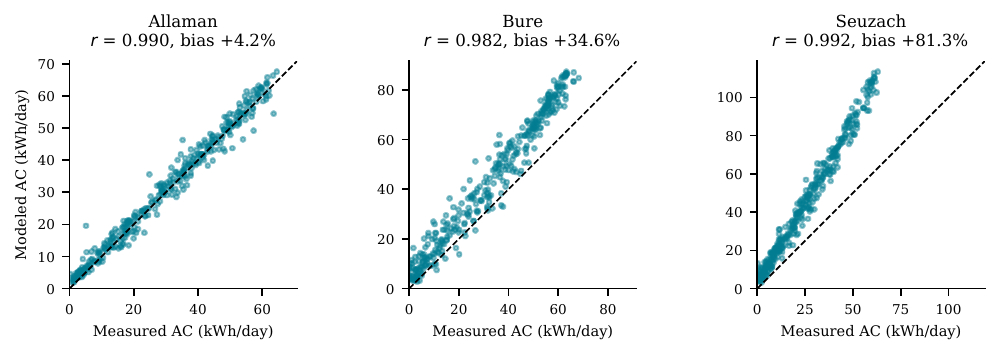}
    \caption{Absolute daily measured and modeled AC energy for 3 installations.
    Each point is a paired day and the dashed line denotes a perfect prediction.
    Titles report daily correlation and signed energy bias.}
    \Description{The three systems show similar measured and modeled daily patterns, but different errors in total energy.}
    \label{fig:daily-examples}
\end{figure}

\parasum{Daily agreement and energy bias.}
We find that the median daily Pearson correlation is \EvalMedianDailyCorrelation{} across the \MeasuredPVOutputCount{} installations.
We also compute correlation after removing each series' monthly mean, to see how well the model captures daily fluctuations within each month (\Cref{fig:yield-validation}a).
The median correlation after this adjustment is \WithinMonthMedianR{}, with a range of \WithinMonthMinR{}--\WithinMonthMaxR{}.

Total energy is estimated less accurately: median absolute bias is \EvalMedianAbsoluteBias\%, and median signed bias is also \ProductionMedianSignedBias\%, indicating a tendency to overestimate production.
The mean absolute and signed biases are \ProductionMeanAbsoluteBias\% and $+\ProductionMeanBias\%$, respectively, due to some outliers (\Cref{tab:production}).

The daily comparisons in \Cref{fig:daily-examples} show that modeled yield can track daily production changes while differing in magnitude.
Estimating absolute production requires accurate information about panel ratings, wiring, and inverter specifications.
Degradation and soiling losses, which our model does not capture, can also affect production.
However, a single ground truth
measurement could correct this bias, since it is a nearly time-invariant linear factor.

\parasum{Effect of additional shading.}
We test whether neighborhood and roof-detail shadows improve agreement, using mean absolute daily error, the average absolute difference between modeled and measured daily energy (\Cref{fig:yield-validation}b).
We annualize this metric by multiplying by 365.
Adding neighborhood and roof-detail shadows reduces this error on \PredictionShadingImproved{} of \MeasuredPVOutputCount{} systems.
The median annualized change is \PredictionMedianAnnualizedMAEChange{}~kWh/year.
Median absolute energy bias decreases from \PredictionBaseMedianBias\% with base-roof shading to \EvalMedianAbsoluteBias\% with full shading.
One possible explanation is that existing panel positions already avoid obvious shading sources, a bias we've also observed in \Cref{sec:rq2-results}.

\parasum{Effect of reported capacity.}
We replace the default capacity of 200~W per square meter of inferred panel area with reported DC capacity, scaling the inverter limit to preserve the DC/AC ratio (\Cref{app:electrical}).
As shown in \Cref{fig:yield-validation}c, median absolute bias falls to \PredictionCapacityMedianBias\% from the capacity information alone, and mean absolute daily error decreases on \PredictionCapacityImproved{} of \MeasuredPVOutputCount{} systems.
The improvement shows that the assumed capacity strongly affects estimated energy.
Nevertheless, factors such as wiring, inverter efficiency, panel degradation, and dirt accumulation remain uncertain and may explain some of the remaining bias.

\rqanswer{rq:yield}{Expected yield captures daily variation (median $r=\EvalMedianDailyCorrelation{}$), but absolute energy remains sensitive to system specifications. Using reported DC capacity reduces median absolute bias from \EvalMedianAbsoluteBias\% to \PredictionCapacityMedianBias\%. Additional shading explains little of the discrepancy.}

\subsection{Shading exposure at panel positions}
\label{sec:rq2-results}

\begin{figure}[!htbp]
    \centering
    \includegraphics[width=0.92\linewidth]{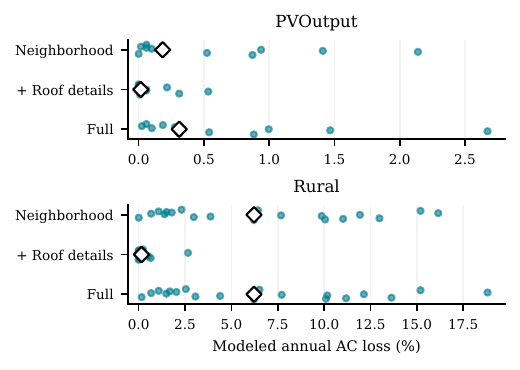}
    \caption{Annual modeled AC losses from neighborhood and roof-detail shading.
    The \MeasuredPVOutputCount{} PVOutput roofs use inferred panels. The \PatchRoofCount{} rural roofs use layouts optimized under base-roof shading at 50\% of the feasible panel count.
    Neighborhood adds surrounding shadows to base-roof shading. + Roof details shows the additional loss when roof-detail shading is added.
    Both increments are percentages of annual AC energy under base-roof shading and add to the total labeled Full.
    Dots show buildings and diamonds show medians. Note that the two datasets use different scales.}
    \Description{Loss under full shading relative to base-roof shading has a median of \FixedMeasuredCombinedMedian{} percent on the PVOutput roofs and \FixedPatchCombinedMedian{} percent on rural layouts designed under base-roof shading. Nearby buildings and trees cause most losses on the rural roofs.}
    \label{fig:fixed-shading}
\end{figure}

The first research question assessed agreement with measurements. We now use the model to isolate the effect of shading at fixed panel positions.
We use inferred layouts for installations with production data and layouts designed under base-roof shading at 50\% of the feasible panel count on rural roofs.
\Cref{fig:fixed-shading} compares annual energy losses in the two settings, with losses normalized by the energy under base-roof shading.

\begin{figure}[!htbp]
    \centering
    \includegraphics[width=0.90\linewidth]{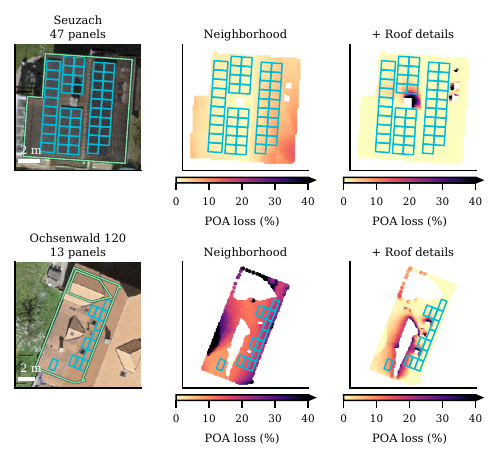}
    \caption{Seuzach's inferred installation (top) and Ochsenwald~120's layout designed under base-roof shading (bottom).
    Left: orthophotos. Middle: neighborhood shading loss. Right: additional loss when roof-detail shading is added.
    Green outlines mark model roof envelopes on orthophotos, and cyan outlines mark PV panels. Losses at each roof position are percentages of annual POA irradiation under base-roof shading at that position.}
    \Description{At Seuzach, roof details cast their strongest shadows mainly between groups of panels. At Ochsenwald, more panels occupy shaded areas. Each row uses the same panel positions.}
    \label{fig:shading-examples}
\end{figure}

\parasum{Losses at existing installations.}
Full shading reduces annual modeled AC energy relative to base-roof shading by a median \FixedMeasuredCombinedMedian\% and at most \FixedMeasuredCombinedMax\%, which is admittedly small, but most likely due to the bias in the choice of homes with publicly-available measurements.
At Seuzach, the installation with the largest relative loss among the PVOutput systems, neighborhood shading removes \FixedSeuzachNeighborhoodLoss\% of energy under base-roof shading and roof details remove a further \FixedSeuzachDetailsLoss\%.
Across the \MeasuredPVOutputCount{} systems, the median neighborhood-level loss and detail increment are negligible.

\parasum{Losses on rural roofs.}
At 50\% of the feasible panel count, loss under full shading on the \PatchRoofCount{} layouts designed under base-roof shading has a median of \FixedPatchCombinedMedian\% and reaches \FixedPatchCombinedMax\% at Ochsenwald~120.
The median neighborhood loss is \FixedPatchNeighborhoodMedian\%.
The largest detail increment is \FixedPatchDetailsMax\% at Ochsenwald~120, while the median is small at \FixedPatchDetailsMedian\%.
Many roof models in this rural set contain no roof details.

The modeled effect therefore depends on the details represented in the input geometry, and missing details would leave their shadows unmodeled.
Where roof details are present, their effect on panel shading can still be small, e.g. if they are not voluminuous, they will not cast significant shadows.

\parasum{Where the losses occur.}
\Cref{fig:shading-examples} shows the overlap between shadows and panels.
At Seuzach, in the existing installation, roof details cast their strongest shadows mainly between groups of panels.
At Ochsenwald~120, more panels in the layout designed under base-roof shading occupy areas affected by both shadow sources.
Visible shading over a roof therefore need not imply a large energy loss at its panel positions, particularly when shadows occur during hours with little sunlight.

The two groups also differ in how their panel positions were selected. Existing installations may already avoid obvious shade, while the rural layouts were generated without neighborhood or roof-detail shadows.

\rqanswer{rq:shading}{Adding neighborhood and roof-detail shading reduces modeled yield by a median \FixedMeasuredCombinedMedian\% on existing installations and \FixedPatchCombinedMedian\% on rural layouts designed under base-roof shading. Neighborhood shadows account for most of this modeled yield reduction on rural roofs.}
\subsection{Relocation and shading-aware new layouts}
\label{sec:rq3-results}

The preceding losses describe sunlight blocked at fixed panel positions. Recovering that energy requires feasible positions with better exposure.
For \Cref{rq:design}, we first test relocation of existing panels, then compare rural layouts generated with different shading information to isolate the benefit of using that information during design.

\begin{figure}[!htbp]
    \centering
    \includegraphics[width=0.86\linewidth]{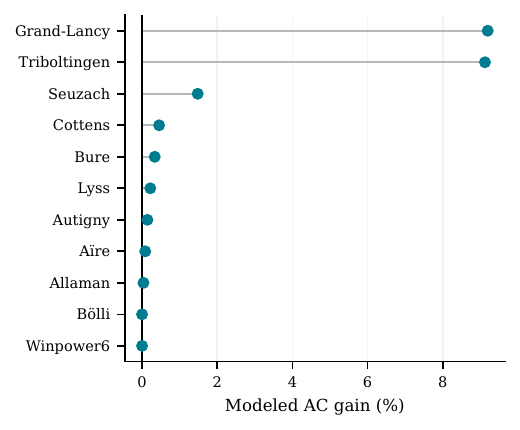}
    \caption{Modeled annual AC gain from relocation on the \MeasuredPVOutputCount{} measured installations, relative to their inferred layouts. Panel counts are unchanged and both layouts are evaluated under full shading.}
    \Description{Relocation gains for all measured installations, ordered from largest to smallest.}
    \label{fig:pvoutput-designs}
\end{figure}

\parasum{Relocation of existing installations.}
Relocation improves modeled AC yield on \DesignRelocationPositive{} of \DesignRelocationFeasible{} roofs.
The median gain is \DesignRelocationMedian\%, with the largest gains at \DesignRelocationMaxHouse{} (\DesignRelocationMax\%) and Triboltingen (\DesignTriboltingenRelocation\%) (\Cref{fig:pvoutput-designs}).
For the other \DesignRelocationRetained{} installations, the search found no layout with higher modeled AC yield.

\begin{figure}[!htbp]
    \centering
    \includegraphics[width=\linewidth]{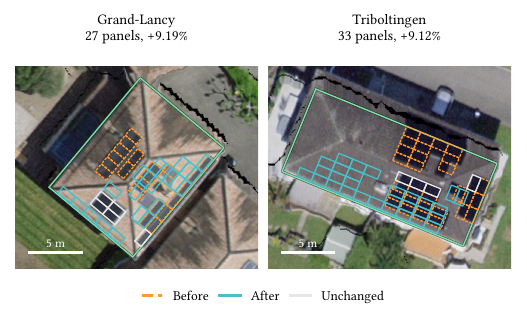}
    \caption{Relocation on Grand-Lancy and Triboltingen. Orange dashed outlines show vacated panel positions, cyan shows relocated positions, and light gray shows unchanged panels. Green marks the model roof envelope. Image artifacts are due to orthorectification.}
    \Description{Corrected orthophotos show inferred and relocated panel arrangements at the two measured installations with the largest modeled relocation gains. Roof envelopes and five-metre scale bars identify the modeled extent and image scale.}
    \label{fig:relocation-examples}
\end{figure}

\Cref{fig:relocation-examples} shows the inferred and relocated panels for Grand-Lancy and Triboltingen, the two installations with the largest modeled gains.
At both, relocation shifts panels toward southern roof segments, increasing modeled yield by \DesignRelocationMax\% and \DesignTriboltingenRelocation\%, respectively.

\parasum{New layouts on rural roofs.}
We compare rural layouts designed under base-roof shading with layouts designed under full shading, both at 50\% of the common feasible panel count.
Both are evaluated under full shading, so the reported gain comes from their different panel positions.
Many panels change position or orientation even though the median energy gain is small.
Using neighborhood and roof-detail shadows moves or rotates a mean \PatchCombinedChangedMean\% of panels per roof across the \PatchRoofCount{} rural roofs.
Compared with layouts designed under base-roof shading, these layouts increase annual modeled AC energy by a median \ResultHalfMedianGain\%, a mean \PatchCombinedGainMean\%, and up to \PatchCombinedGainMax\%.
Only \ResultHalfAboveOne{} of \PatchRoofCount{} roofs gain more than 1\%.
PV layouts for all roofs are provided in \Cref{app:patch}.
\Cref{tab:design-ablations} separates the two shading contributions on the rural roofs.
Accounting for nearby buildings and trees during placement provides most of the average gain, at \PatchNeighborhoodGainMean\%.
Using roof-detail shadows during placement gives a minimal increment on average.
Despite the small gains, neighborhood-aware and full layouts are identical on only \ResultLocalUnchanged{} of \PatchRoofCount{} roofs.

\begin{table}[!htbp]
    \centering\small
    \caption{Effect of the shading sources used during design on new layouts for the \PatchRoofCount{} rural roofs at 50\% of the common feasible panel count.
    A / B compares layouts designed under settings A and B, both evaluated under full shading. Gain is relative to B. Changed is the percentage of panels moved or rotated.}
    \label{tab:design-ablations}
    \input{artifacts/rq_design_ablation_table}
\end{table}

\begin{figure}[!htbp]
    \centering
    \includegraphics[width=0.92\linewidth]{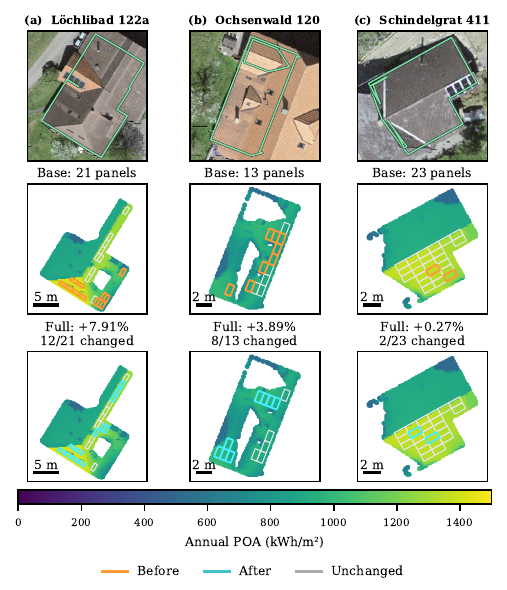}
    \caption{Three rural roofs at 50\% of the feasible panel count: orthophotos (top), layouts designed under base-roof shading (middle) and full shading (bottom).
    Orange and cyan mark changed panel positions before and after redesign, respectively. Gray marks unchanged panels.
    The background color scale shows annual POA irradiation under full shading. AC gain and changed-panel count are labeled.}
    \Description{Three before-and-after examples highlight the actual changed panels against full-scene irradiation. Every column compares the same two layout types at equal count.}
    \label{fig:patch-layout}
\end{figure}

\parasum{Examples of layout changes.}
\Cref{fig:patch-layout} shows (a) a larger placement benefit, (b) a contribution from roof-detail shading, and (c) a small yield benefit with few changed panels. Each example compares layouts designed under base-roof and full shading.

At Löchlibad~122a (a), \ResultExampleAChanged{} of \ResultExampleACount{} panels change position or orientation.
This adds \ResultExampleAKWh{}~kWh/year, a \ResultExampleAGain\% gain.

At Ochsenwald~120 (b), the full-shading layout changes \ResultExampleBChanged{} of \ResultExampleBCount{} panels.
The annual gain over the layout designed under base-roof shading is \ResultExampleBKWh{}~kWh (\ResultExampleBGain\%).
Including roof-detail shadows changes \PatchExampleChanged{} panels relative to the neighborhood-aware layout and adds \PatchExampleGainKWh{}~kWh/year.
This is a \PatchLocalGainMax\% gain relative to the neighborhood-aware layout.

At Schindelgrat~411 (c), changing \ResultExampleCChanged{} of \ResultExampleCCount{} panels yields a minimal gain of \ResultExampleCGain\%, or \ResultExampleCKWh{}~kWh/year.
The layouts cover largely the same roof area, with \ResultExampleCShared\% shared panel area.

\begin{figure*}[!tbp]
    \centering
    \includegraphics[width=0.99\linewidth]{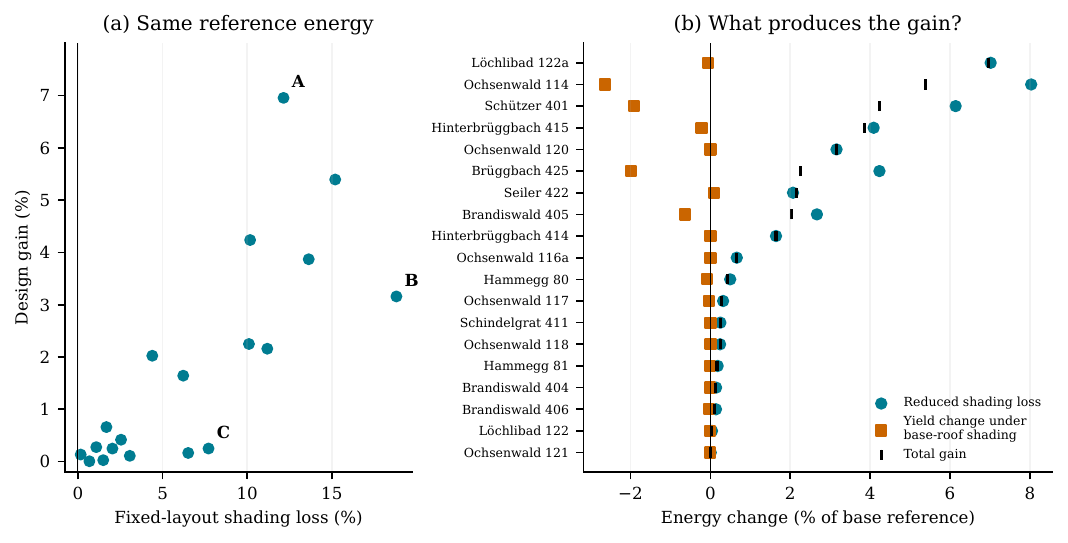}
    \caption{Shading exposure and placement benefit on all \PatchRoofCount{} rural roofs at 50\% of the feasible panel count.
    (a) Loss from adding neighborhood and roof-detail shading to a layout designed under base-roof shading, versus the gain from redesigning it under full shading.
    (b) The gain decomposes into reduced shading loss and a yield change under base-roof shading, as derived in \Cref{eq:energy-accounting}.
    The two colored values sum to the total gain for each roof.
    All percentages are normalized by the energy of $\Layout{Base}$ evaluated under base-roof shading.
    A--C correspond to (a)--(c) in \Cref{fig:patch-layout}.}
    \Description{A matched comparison links shading exposure to placement benefit and separates each gain into avoided shading loss and a yield change under base-roof shading. Most layouts designed under full shading receive less energy under base-roof shading but avoid a larger shading penalty.}
    \label{fig:energy-accounting}
\end{figure*}

\parasum{Sources of the energy gain.}
A layout can avoid more shade yet give up some sunlight available from a better-oriented roof segment.
\Cref{fig:energy-accounting} separates these effects using \Cref{eq:energy-accounting}: the energy change under base shading and the neighborhood and roof-detail loss avoided by moving panels.
This decomposition uses a different percentage denominator from the primary design-gain results. Both components are divided by the energy of $\Layout{Base}$ under base-roof shading, so they sum to the total shown.
The primary design gains divide by the energy of the same layout under full shading.

All revised layouts on rural roofs reduce the additional shading penalty.
On \EnergyBaselineNegative{} of \PatchRoofCount{} roofs, the new positions receive less energy under base shading but avoid a larger loss from the additional shadows.

Ochsenwald~120 has the largest shading loss on a layout designed under base-roof shading (\MatchedBLoss\%), yet gains less from designing under full shading than Löchlibad~122a.
The most shaded roof therefore need not achieve the largest yield gain from changing panel positions.


\parasum{Effect of panel count.}
More panels leave fewer alternative positions to avoid shade. Across the rural roofs, mean relative gain falls from \BudgetLowGainMean\% at 20\% of the feasible count to \BudgetFullGainMean\% at full capacity, although individual roofs need not follow this trend. Mean additional energy peaks at 50\% of the feasible count. \Cref{app:panel-count} gives the full comparison. These results do not determine the best installation size.

\rqanswer{rq:design}{Relocation improves modeled yield at \DesignRelocationPositive{} of \DesignRelocationFeasible{} existing installations. Rural shading-aware design gains a median \ResultHalfMedianGain\% and up to \PatchCombinedGainMax\% at 50\% of the feasible panel count, mainly by avoiding neighborhood shadows. Benefits vary with roof geometry and panel count. These modeled energy gains remain to be validated on redesigned installations.}

%% file: artifacts/production_table.tex
\begin{tabular}{@{}lrrrrrrr@{}}
\toprule
House & Year & Panels & Days & \shortstack{Measured AC\\{}[MWh]} & \shortstack{Modeled AC\\{}[MWh]} & Bias [\%] & $r$ \\
\midrule
Aïre & 2023 & 11/9 & 351 & 1.76 & 6.15 & +248.5 & 0.911 \\
Allaman & 2023 & 24/24 & 363 & 10.92 & 11.38 & +4.2 & 0.990 \\
Autigny & 2023 & 30/44 & 364 & 14.13 & 14.62 & +3.5 & 0.950 \\
Bure & 2023 & 42/34 & 364 & 11.91 & 16.03 & +34.6 & 0.982 \\
Cottens & 2023 & 34/35 & 357 & 11.58 & 15.53 & +34.2 & 0.983 \\
Grand-Lancy & 2023 & 27/20 & 364 & 5.14 & 9.76 & +90.0 & 0.972 \\
Bölli & 2023 & 25/35 & 364 & 12.19 & 9.18 & -24.7 & 0.977 \\
Lyss & 2019 & 42/32 & 364 & 8.35 & 20.42 & +144.6 & 0.954 \\
Seuzach & 2023 & 47/48 & 364 & 9.25 & 16.77 & +81.3 & 0.992 \\
Triboltingen & 2023 & 33/25 & 364 & 10.83 & 11.28 & +4.2 & 0.990 \\
Winpower6 & 2023 & 20/22 & 364 & 5.93 & 8.34 & +40.8 & 0.985 \\
\bottomrule
\end{tabular}

%% file: artifacts/rq_design_ablation_table.tex
\begin{tabular}{@{}lrrr@{}}
\toprule
& \multicolumn{2}{c}{AC gain [\%]} & Changed [\%] \\
Design comparison & Mean & Median & Mean \\
\midrule
Neighborhood / base & 1.947 & 0.671 & 35.2 \\
Full / base & 2.027 & 0.671 & 38.9 \\
Full / neighborhood & 0.079 & 0.000 & 11.1 \\
\bottomrule
\end{tabular}

%% file: sections/discussion.tex
\section{Discussion}
\label{sec:discussion}

Our model captures daily production changes, however absolute yield remains sensitive to system specifications. Shading loss alone does not predict redesign gains, which depend on available roof space and panel count. On the rural roofs, neighborhood shadows dominate both losses and design gains. Roof-detail shadows have a smaller average effect but can still change panel positions.

Relocation can improve an inferred installation for several reasons, including moves toward better-oriented roof segments, as seen at Grand-Lancy and Triboltingen.
The rural comparisons isolate the effect of the shading information used during design.
Because the placement procedure and constraints remain the same, we can attribute their yield differences to the use of neighborhood and roof-detail shadows.

\subsection{Limitations}
Segmentation errors and imagery misalignment can distort inferred panel counts and positions. Missing roof details leave obstructions and shadows unmodeled, affecting yield and placement estimates. Production measurements allow comparison only for existing installations. Redesign gains are thus not measured, but merely modeled in simulation. That being said, our simulation is deterministic and allows comparing relative changes in yield under fixed electrical assumptions.

\balance

The inferred existing layouts have small modeled shading losses, we believe due to selection bias,
and the degree of shading from surrounding buildings is limited in the studied settings.
Dense urban areas with tall buildings would offer a useful test of placement under stronger building shadows.
The evaluation also covers only Swiss houses, so it does not establish how the findings generalize to other building types and climates.

The shading model treats vegetation as opaque and applies obstruction shadows only to direct sunlight.
It omits seasonal canopy transmission, obstruction of diffuse light, mutual panel shading, and tilted panel modeling on flat roofs.
It does not establish a globally optimal layout or account for all maintenance clearance, wiring, and structural requirements.

\subsection{Future work}
Surveyed installations with precisely known system and condition specifications, including shaded systems, would let us examine geometric and electrical errors separately.
Panel-level measurements could then test localized shading effects and inform string-aware modeling.

The workflow could extend to other regions with suitable geometry, imagery, LiDAR, and weather data. Adding installation constraints and costs would help assess whether redesign is worthwhile. Joint optimization of panel count and placement could support sizing decisions~\citep{gschwind_joint_2026}.

%% file: sections/conclusion.tex
\section{Conclusion}

We presented a workflow and public tool that combines Sun-view shadow rendering with hourly weather to estimate PV yield, relocate inferred panels, and design new layouts with terrain, neighborhood, and roof-detail shading.

Daily production agrees closely with modeled yield, but absolute energy remains sensitive to system specifications. Relocation improves modeled yield on \DesignRelocationPositive{} of \MeasuredPVOutputCount{} installations. On \PatchRoofCount{} rural roofs at half the feasible panel count, shading-aware design gains a median \ResultHalfMedianGain\% and up to \PatchCombinedGainMax\%, mainly by avoiding neighborhood shadows. These redesign gains remain to be tested in real installations.

Our public tool lets users inspect shading, assess existing installations, compare relocated layouts, and model new layouts at different panel counts for any building in Switzerland.

%% file: sections/appendix.tex
\section{Implementation details}
\label[appendix]{app:settings}

This appendix specifies the numerical settings used by the workflow in the main text: fitting existing panels, generating candidate positions, sampling sunlight, and converting it to electrical energy.

\parasum{Inferring existing layouts.}
Panel fitting maximizes $0.82F_1+0.18B$, combining agreement with the segmented PV area ($F_1$) and alignment with visible panel boundaries ($B$).
Let $P$ be the image pixels covered by the fitted panels and $M$ the pixels classified as PV. The overlap score $F_1=2|P\cap M|/(|P|+|M|)$ is the harmonic mean of precision and recall. It equals 1 for identical masks and 0 for no overlap, penalizing both missed PV area and coverage outside the segmentation.
The boundary score $B\in[0,1]$ measures image-edge strength along boundaries shared by adjacent fitted panels. It averages normalized grayscale gradients across those boundaries, favoring grids aligned with visible panel divisions. We set $B=0$ when no shared boundaries exist.
The weights are heuristic choices that emphasize segmentation overlap while using internal image edges to distinguish similar panel grids. They were not estimated from labeled layouts.
Among candidates scoring within 0.01 of the best, we prefer dimensions closest to common panel sizes.

\parasum{Placement defaults.}
Grid offsets are tested in 0.10~m steps.
Roof-edge, ridge, and obstruction clearances are 0.05~m, with zero inter-panel gap.
This assumed 5~cm margin shrinks usable roof polygons and expands obstruction footprints. It is not calibrated from data and does not establish installation compliance.
 For sunlight sampling, we use a horizontal grid with 0.20~m spacing and place each receiver 0.05~m above the roof along its surface normal.
The offset represents assumed mounting height above the roof (\Cref{sec:shadow-states}).

\parasum{Comparing layouts.}
After pairing panels between two layouts as described in \Cref{sec:evaluation}, we count a panel as changed if its center moves more than 0.25~m, or its long axis or roof normal rotates more than $5^\circ$.

\parasum{Snow treatment.}
 For the production comparison in \Cref{rq:yield}, we use pvlib's National Renewable Energy Laboratory (NREL) snow model~\citep{pvlib_snow_nrel} with MeteoSwiss observations when a supported station within 25~km provides snow depth, precipitation, and temperature.
The shading and placement comparisons in \Cref{rq:shading,rq:design} omit snow losses.
Snowfall is approximated from positive hourly increases in station snow depth when precipitation occurred within the preceding three hourly records and station air temperature is at most $2\,{}^\circ$C.
Station snow depth describes ground conditions. Panel tilt, facet-mean POA irradiance, and air temperature determine modeled snow coverage and sliding~\citep{pvlib_snow_nrel}.
The model accounts for slope without estimating snow depth on the panels.

\subsection{Electrical model}
\label[appendix]{app:electrical}

Defaults use 200~W of rated DC power per square metre of panel area, equivalent to 20\% efficiency at the standard irradiance of 1000~W/m$^2$.
This is a representative assumption between the 19\% and 21\% efficiencies used for standard and premium crystalline-silicon modules in PVWatts~\citep{sam_pvwatts_system_design}.
We use its nominal inverter efficiency of 96\%, approximated here as constant, and choose a DC/AC ratio of 1.15 within its documented typical range of 1.10--1.25~\citep{sam_pvwatts_system_design}.
These are common assumptions across installations. Electrical mismatch between shaded panels is not modeled.
For panel area $A$ in m$^2$ and count $N$, default rated DC power is $P_0=200NA/1000$~kW.
With hourly array-mean POA $G_t$ in W/m$^2$, the electrical model is
\begin{align}
P_{\mathrm{DC},t} &= P_0\frac{G_t}{1000}
\max\{0,1+\gamma(T_{\mathrm{cell},t}-25)\}\notag\\
&\qquad\times(1-L_{\mathrm{snow},t})\quad[\mathrm{kW}]\\
E_{\mathrm{AC},t} &= \min\{P_0/1.15,\ 0.96P_{\mathrm{DC},t}\}\,\Delta t\quad[\mathrm{kWh}]
\end{align}
 Here, $T_{\mathrm{cell},t}$ is panel temperature in degrees Celsius, $\gamma=-0.0035\,/{}^\circ$C is the temperature coefficient, and $L_{\mathrm{snow},t}$ is the modeled fraction of DC output lost to snow.
We set $L_{\mathrm{snow},t}=0$ for the shading and placement comparisons.
The one-hour interval $\Delta t=1$~h converts power to energy in kWh.
The temperature coefficient equals the PVWatts premium crystalline-silicon value of $-0.35\%/{}^\circ$C~\citep{sam_pvwatts_system_design}.
We estimate panel temperature with the Faiman model~\citep{faiman_temperature_2008}, $T_{\mathrm{cell},t}=T_{\mathrm{air},t}+G_t/(U_0+U_1v_t)$, using PVGIS air temperature and wind speed $v_t$.
We retain pvlib's defaults $U_0=25$~W/(m$^2$\,K) and $U_1=6.84$~W\,s/(m$^3$\,K), derived from measurements of modules on an open rack~\citep{pvlib_faiman_defaults}. Their transfer to roof-mounted panels is an approximation.
No additional general loss factor is applied, leaving wiring, soiling, and availability losses unmodeled.
The capacity diagnostic replaces $P_0$ with the published DC capacity and retains the DC/AC ratio.

\section{Effect of panel count}
\begin{figure}[!htbp]
    \centering
    \includegraphics[width=\linewidth]{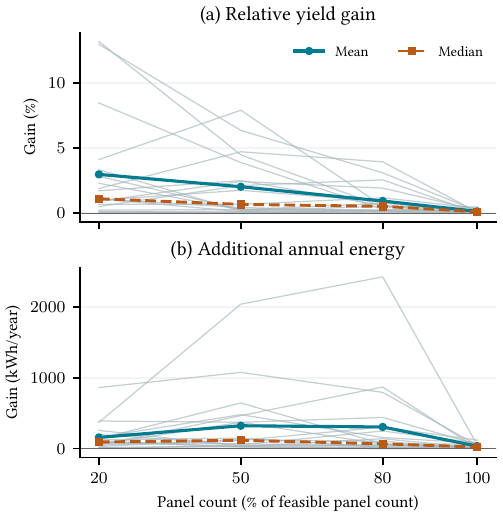}
    \caption{Effect of panel count on relative and absolute placement gains.
    Each thin line is one of the \PatchRoofCount{} rural roofs. Thick lines show the mean and median.
    Both layouts have the same panel count at each fraction of the common feasible panel count and are evaluated under full shading.
    Relative gain is normalized by the AC energy under full shading of the layout designed under base-roof shading.}
    \Description{Two panels show per-roof percentage and annual kilowatt-hour gains at four panel counts. Mean relative gain decreases, while mean extra energy peaks at 50\% of the feasible panel count.}
    \label{fig:budget-response}
\end{figure}

\label[appendix]{app:panel-count}
\Cref{fig:budget-response} shows how placement gains vary with panel count.

The main rural comparisons use half the feasible panel count. Increasing the number of panels leaves fewer alternative positions, which can limit the opportunity to avoid shade.
On the rural roofs, we compare layouts designed with only base-roof shading against those designed with full shading at different panel counts (\Cref{fig:budget-response}).
The mean percentage yield gain from using full shading during design falls from \BudgetLowGainMean\% at 20\% of the feasible panel count to \BudgetFullGainMean\% at the full feasible panel count.
The median falls from \ResultLowMedianGain\% to \ResultFullMedianGain\%.
Some roofs show increases in gain between tested panel counts, as added panels can have different shading exposure in the two designs.
Mean additional yield rises from \ResultLowMeanKWh{}~kWh/year at 20\% to \ResultHalfMeanKWh{}~kWh/year at 50\% of the feasible panel count, then falls to \ResultHighMeanKWh{}~kWh/year at 80\% and \ResultFullMeanKWh{}~kWh/year at the full feasible panel count.
Percentage gain is largest at low panel counts, while mean additional yield peaks at 50\% of the feasible panel count on these rural roofs.
These comparisons show how panel count affects the benefit of shading-aware design, without determining the best installation size.

\section{Interface}
\begin{figure*}[!tbp]
\centering
\begin{minipage}{0.85\textwidth}
\noindent
\begin{minipage}[t]{0.49\linewidth}
  \centering
  \begin{minipage}[c][0.72\linewidth][c]{\linewidth}
    \centering
    \includegraphics[width=\linewidth,height=0.72\linewidth,keepaspectratio]{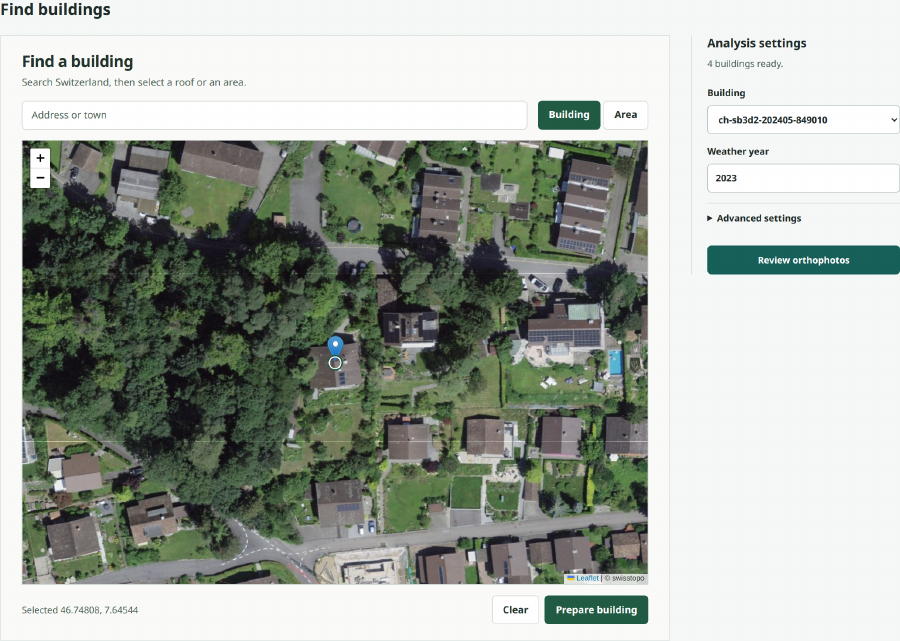}
  \end{minipage}\par
  \vspace{1mm}
  (a) Find a building
\end{minipage}%
\hfill
\begin{minipage}[t]{0.49\linewidth}
  \centering
  \begin{minipage}[c][0.72\linewidth][c]{\linewidth}
    \centering
    \includegraphics[width=\linewidth,height=0.72\linewidth,keepaspectratio]{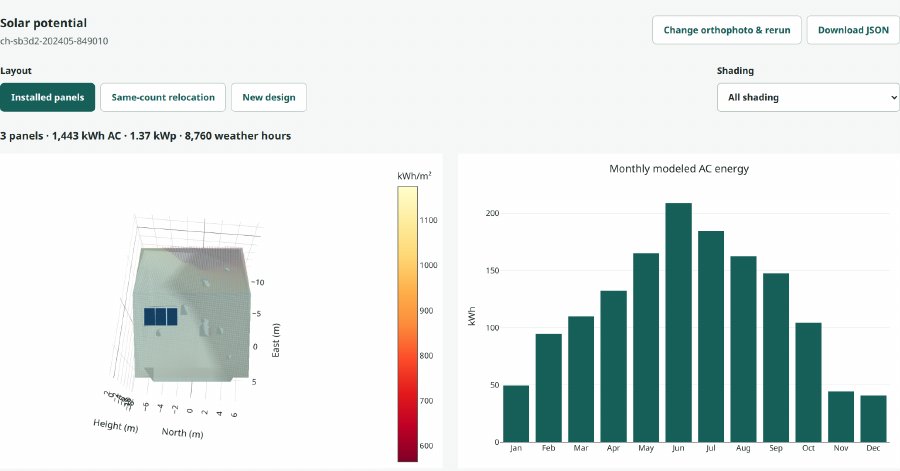}
  \end{minipage}\par
  \vspace{1mm}
  (b) Inspect existing panels
\end{minipage}

\vspace{2mm}
\noindent
\begin{minipage}[t]{0.49\linewidth}
  \centering
  \begin{minipage}[c][0.72\linewidth][c]{\linewidth}
    \centering
    \includegraphics[width=\linewidth,height=0.72\linewidth,keepaspectratio]{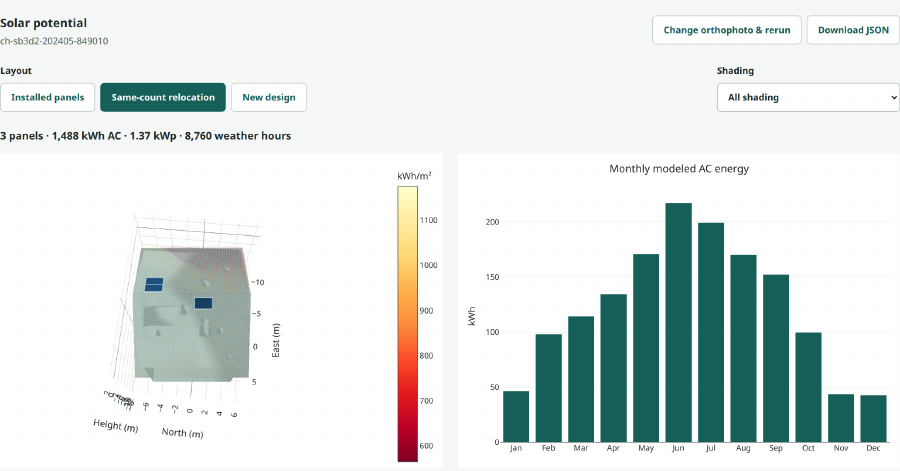}
  \end{minipage}\par
  \vspace{1mm}
  (c) Compare relocation
\end{minipage}%
\hfill
\begin{minipage}[t]{0.49\linewidth}
  \centering
  \begin{minipage}[c][0.72\linewidth][c]{\linewidth}
    \centering
    \includegraphics[width=\linewidth,height=0.72\linewidth,keepaspectratio]{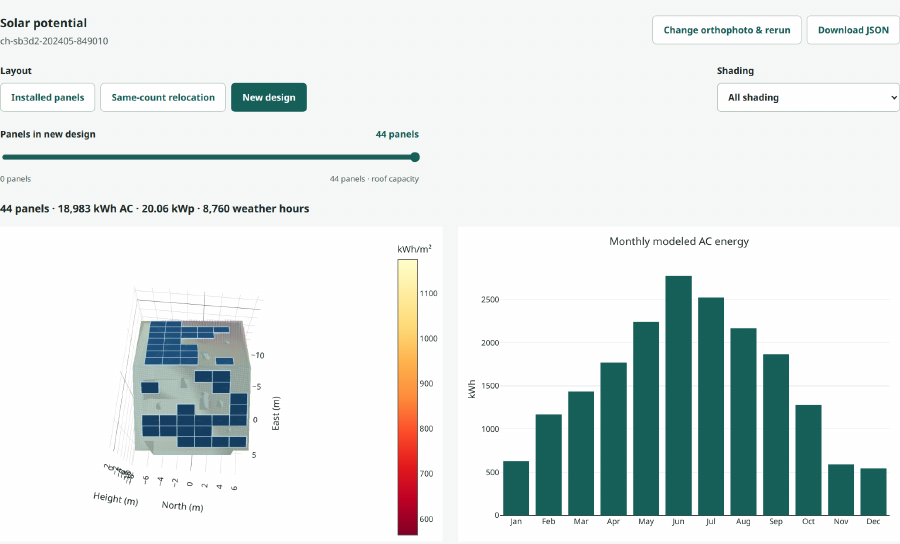}
  \end{minipage}\par
  \vspace{1mm}
  (d) Adjust the panel count
\end{minipage}
\end{minipage}
\caption{Web interface: (a) select a building on the map or via an address, (b) inspect inferred panels and expected yield, (c) compare relocation benefits, and (d) explore new layouts at variable panel counts.}
\Description{Four application screenshots show map-based building selection, an inferred installation with production estimates, a relocated layout, and the panel-count slider for new layouts.}
\label{fig:shading-aware-pv-workflow}
\end{figure*}

\label[appendix]{app:interface}
\Cref{fig:shading-aware-pv-workflow} shows how users inspect existing installations and compare alternative layouts.

\section{Complete results and layouts}
\label[appendix]{app:patch}
\label[appendix]{app:house-atlas}

\begin{figure*}[!tbp]
    \centering
    \includegraphics[width=0.95\textwidth]{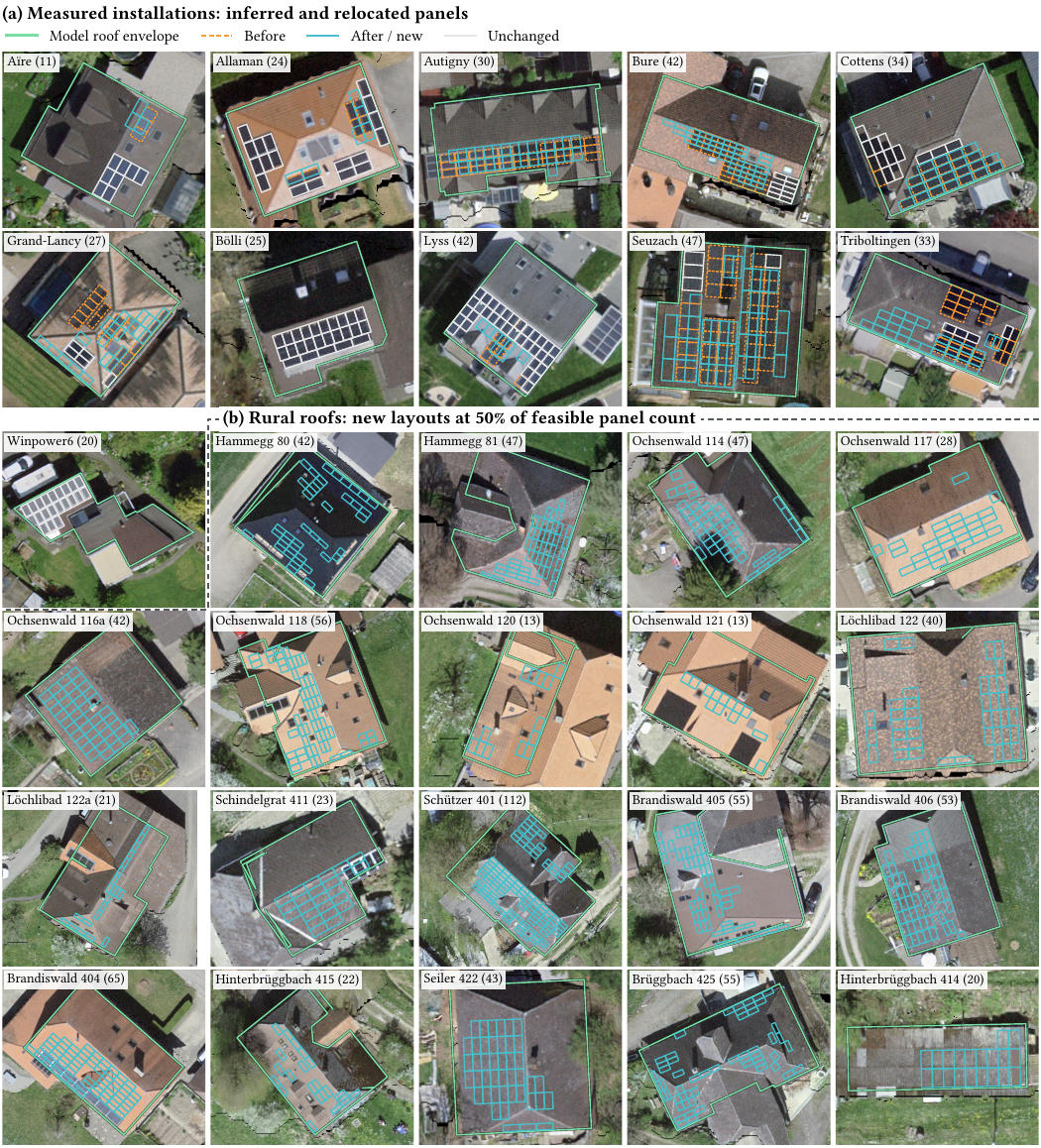}
    \caption{All evaluated roofs. Panel counts are indicated in parentheses.
    The divider separates (a) \MeasuredPVOutputCount{} measured installations and (b) \PatchRoofCount{} rural roofs.
    (a) Orange dashed outlines show vacated panel positions, cyan shows relocated positions, and light gray shows unchanged panels. Panel count is unchanged.
    (b) Cyan shows full-shading layouts at 50\% of the common feasible panel count.}
    \Description{A tightly spaced grid of five columns and six rows shows thirty corrected orthophotos. Green outlines identify model roof envelopes. The first eleven houses show vacated positions in orange, relocated positions in cyan, and unchanged panels in light gray. The remaining nineteen have new full-shading layouts at 50\% of the feasible panel count. A stepped dashed divider separates the two groups.}
    \label{fig:house-atlas}
\end{figure*}

\Cref{tab:cohort-results} reports all \PatchRoofCount{} rural buildings, including small and negative increments.
\Cref{fig:house-atlas} shows every evaluated roof with its imagery, model envelope, and panel layouts.

\begin{table*}[!tbp]
    \centering\footnotesize
    \caption{Placement results for all \PatchRoofCount{} rural roofs, evaluated under full shading.
    $N$ is 50\% of the feasible panel count. Neigh., Local, and Full give relative yield gains for neighborhood/base, full/neighborhood, and full/base designs.
    Changed reports panels moved or rotated from base to full shading.
    The final columns give full/base gains at 20\% and 100\% of the feasible panel count.}
    \label{tab:cohort-results}
    \input{artifacts/rq_cohort_table}
\end{table*}

%% file: artifacts/rq_cohort_table.tex
\begin{tabular}{@{}lrrrrrrr@{}}
\toprule
& & \multicolumn{4}{c}{50\% of feasible panel count} & \multicolumn{2}{c}{Full/base gain [\%]} \\
\cmidrule(lr){3-6}\cmidrule(l){7-8}
Building & $N$ & Neigh. [\%] & Local [\%] & Full [\%] & Changed [\%] & 20\% & 100\% \\
\midrule
Hammegg 80 & 42 & 0.379 & 0.050 & 0.429 & 7.1 & 1.080 & 0.163 \\
Hammegg 81 & 47 & 0.177 & 0.000 & 0.177 & 4.3 & 3.338 & 0.029 \\
Ochsenwald 114 & 47 & 6.359 & 0.000 & 6.359 & 34.0 & 12.984 & 0.156 \\
Ochsenwald 117 & 28 & 0.279 & -0.000 & 0.279 & 7.1 & 0.227 & 0.177 \\
Ochsenwald 116a & 42 & 0.671 & 0.000 & 0.671 & 100.0 & 0.671 & 0.382 \\
Ochsenwald 118 & 56 & 0.247 & 0.007 & 0.254 & 64.3 & 2.294 & 0.201 \\
Ochsenwald 120 & 13 & 3.231 & 0.637 & 3.889 & 61.5 & 8.468 & 0.105 \\
Ochsenwald 121 & 13 & 0.008 & -0.001 & 0.007 & 15.4 & 0.032 & 0.219 \\
Löchlibad 122 & 40 & 0.034 & -0.006 & 0.028 & 10.0 & 0.132 & 0.007 \\
Löchlibad 122a & 21 & 7.915 & 0.000 & 7.915 & 57.1 & 4.113 & -0.017 \\
Schindelgrat 411 & 23 & 0.271 & 0.000 & 0.271 & 8.7 & 2.831 & 0.025 \\
Schützer 401 & 112 & 4.720 & -0.004 & 4.716 & 18.8 & 1.886 & 0.069 \\
Brandiswald 405 & 55 & 1.985 & 0.129 & 2.117 & 16.4 & 0.505 & 0.002 \\
Brandiswald 406 & 53 & 0.114 & 0.000 & 0.114 & 5.7 & 1.103 & 0.023 \\
Brandiswald 404 & 65 & 0.003 & 0.132 & 0.135 & 100.0 & 0.078 & 0.018 \\
Hinterbrüggbach 415 & 22 & 4.479 & 0.000 & 4.479 & 50.0 & 13.204 & 0.462 \\
Seiler 422 & 43 & 2.430 & 0.000 & 2.430 & 93.0 & 0.913 & 0.228 \\
Brüggbach 425 & 55 & 1.943 & 0.548 & 2.501 & 56.4 & 1.715 & 0.236 \\
Hinterbrüggbach 414 & 20 & 1.752 & 0.000 & 1.752 & 30.0 & 0.991 & 0.000 \\
\bottomrule
\end{tabular}